\documentclass[12pt,preprint]{aastex}
\newcommand{\msun}      {\ensuremath{M_\odot}}
\newcommand{\kms}       {\ensuremath{~\mathrm{km~s^{-1}}}}
\newcommand{\au}        {\ensuremath{~\mathrm{AU}}}

\begin{document}

\title{Growth of Intermediate-Mass Black Holes in Globular Clusters}
\author{Kayhan G\"{u}ltekin} \author{M. Coleman Miller}
\author{Douglas P. Hamilton} \affil{University of Maryland, College
  Park, Dept.\ of Astronomy}

\begin{abstract}
  We present results of numerical simulations of sequences of
  binary-single scattering events of black holes in dense stellar
  environments.  The simulations cover a wide range of mass ratios
  from equal mass objects to 1000:10:10$~\msun$ and compare purely
  Newtonian simulations to simulations in which Newtonian encounters
  are interspersed with gravitational wave emission from the binary.
  In both cases, the sequence is terminated when the binary's merger
  time due to gravitational radiation is less than the arrival time of
  the next interloper.  We find that black hole binaries typically
  merge with a very high eccentricity ($0.93 \le e \le 0.95$ pure
  Newtonian; $0.85 \le e \le 0.90$ with gravitational wave emission)
  and that adding gravitational wave emission decreases the time to
  harden a binary until merger by $\sim30$ to $40\%$.  We discuss the
  implications of this work for the formation of intermediate-mass
  black holes and gravitational wave detection.

\end{abstract}


\section{Introduction}

Recent observations suggest that large black holes may reside in the
centers of some stellar clusters.  X-ray observations in the last few
years have shown unresolved sources in galaxies offset from their
nuclei and with fluxes that, if isotropic, correspond to luminosities
of $L \approx 10^{39}$ to $10^{41}~{\mathrm {erg}}\; {\mathrm s}^{-1}$
\citep*[e.g.,][]{fsm97, cm99, metal01, fzm01}.  Many of these sources
are associated with stellar clusters (\citealt{fsm97};
\citealt*{alm01}).  The strong variability observed in these sources
suggests that they are black holes, and if the observed fluxes are
neither strongly beamed nor super-Eddington, the implied masses are as
high as $M\ga 10^{3}\;{M}_{\odot}$.  The fact the sources are
non-nuclear implies masses $M\la 10^{6}\;{M}_{\odot}$ since a larger
mass would have rapidly sunk to the center of the host galaxy due to
dynamical friction \citep[$< 10^{9}$~yr for a dispersion velocity of
$100\kms$ and a separation from the galaxy nucleus of
$10^{2}~\mathrm{pc}$ as in the case of M82;][]{ketal01}.  In addition, 
optical observations of the globular clusters M15 and G1 show velocity
profiles consistent with central black holes with masses of $2.5\times
10^{3}~\msun$ and $2.0\times 10^{4}~\msun$, respectively
\citep*{getal00, getal02, vetal02, grh02}, although \citet{betal03}
demonstrate with their N-body simulations that the observations of G1
can be explained without a large black hole.  Such intermediate-mass
black holes (IMBHs) would be in a different mass category, and thus
likely indicative of a different formation scenario, from either $3$ -
$20~M_{\odot}$ stellar-mass black holes, which are thought to be the
result of core-collapse supernovae, or $10^{6}$ - $10^{10}~M_{\odot}$
supermassive black holes, which are found in the centers of many
galaxies.

Several models have been proposed to account for the origin of IMBHs.
\citet{mr01} and \citet{setal02} suggest that they are the remnants of
massive ($M \ga 200~\msun$) Population~III stars.  The low metallicity
of these stars precludes cooling through metal line emission and
enables them to reach masses much larger than ordinary main sequence
stars.  These large stars avoid significant mass loss due to stellar
winds or pulsations, and the star may collapse to form a black hole
with almost the same mass as the progenitor star.  \citet{pzm02} and
\citet*{gfr04} show with numerical simulations that the core of a young
stellar cluster may collapse rapidly such that direct collisions of
stars will lead to runaway growth of a single object with as much as
$10^{-3}$ of the original cluster mass over the course of a few
million years.  \citet{mh02} propose that over a Hubble time
stellar-mass black holes in dense globular clusters may grow by
mergers to the inferred IMBH masses.  In their model, a black hole
with mass greater than $50~\msun$ will interact with other massive
objects to form binaries that will merge due to gravitational
radiation.  The merger process may proceed more quickly in the
presence of encounters with a third black hole or another black hole
binary \citep{mh02b} if the encounters shrink the binary's orbit, as
is known to happen with hard (tight) binaries \citep{h75}.

Wherever and however IMBHs formed, the best candidates are found in
stellar clusters where three-body encounters are important.  An IMBH
in a cluster, whether formed there or later swallowed by the cluster,
will find its way to the center.  As all of the heaviest objects in a
cluster sink to the center in a process known as mass segregation, the
IMBH will interact primarily with other massive objects and binaries
\citep{sh93, fetal02}.  A single IMBH will tend to acquire companions
through exchanges with binaries because the most massive pair of
objects in a three-body encounter preferentially end up in the binary
\citep*[e.g.,][]{hhm96}.  The IMBH binary will encounter other objects
in the dense center of its host cluster, harden further, and
ultimately merge.

These events are important sources of gravitational waves.  The
Advanced LIGO (Laser Interferometer Gravitational Wave Observatory)
detector is expected to be capable of detecting mergers of IMBHs with
$M \la 100~\msun$ \citep*{b00}, and LISA (Laser Interferometer Space
Antenna) is expected to detect the earlier inspiral phase of an IMBH
merger \citep*{d00}.  In order to predict the gravitational wave
signature of the inspiral, the expected separations and eccentricities
of the binaries must be known.  Because three-body encounters alter
the orbital parameters, simulations are needed to predict their
distributions as well as the source population and event rates.

The three-body problem has been studied extensively, but with every
new generation of computing power, our understanding of the problem
advances with a wider range of numerical simulations and a changing
perspective on this rich but conceptually simple problem.  Previous
studies of the three-body problem have tended to focus on the case of
equal or nearly equal masses \citep[e.g.,][]{h75, hb83} though other
mass ratios have been studied \citep[e.g.,][]{fh82, sp93, hhm96}.  The
nearly equal mass case does not apply to the case of an IMBH in the
core of a stellar cluster.  In addition the vast majority of previous
work has studied the effect of a single encounter on a binary.  To
determine the ultimate fate of an IMBH, simulations of sequences of
encounters are needed. Furthermore, to our knowledge no previous work
has considered the effects of orbital decay due to gravitational
radiation between encounters, which we expect to be important for very
tight binaries.

In this paper we present numerical simulations of sequences of
high-mass ratio binary-single encounters.  We describe the code used
to simulate the encounters in \S~\ref{nummeth}.  Next, we present
results of the simulations of sequences of encounters on a range of
mass ratios with Newtonian gravity (\S~\ref{newt}) and with
gravitational radiation between encounters (\S~\ref{grseq}) and show
that including gravitational radiation decreases the duration of the
sequence by $\sim 30$ to $40\%$.  In \S~\ref{imbhformation} and
\S~\ref{gwdetection} we discuss the implications of these results for
IMBH formation and gravitational wave detection.

\section{Numerical Method}
\label{nummeth}
We perform numerical simulations of the interactions of a massive
binary in a stellar cluster.  Simulating the full cluster is beyond
current N-body techniques, so we focus instead on a sequence of
three-body encounters.  Massive cluster objects, such as IMBHs and 
tight binary systems, tend to sink the centers of clusters so that a
single IMBH is very likely to meet a binary \citep{sp95}.  Exchanges
in which the IMBH acquires a close companion are common.  Such a
binary in a stellar cluster core will experience repeated interactions
with additional objects as long as the recoils from these interactions
do not eject the binary.  Therefore, we simulate a \emph{sequence} of
encounters between a hard binary and an interloper.  We perform one
interaction and then use the resulting binary for the next encounter.
This is repeated multiple times until the binary finally merges due to
gravitational radiation.  Because typical velocities involved are
non-relativistic and the black holes are tiny compared to their
separations, they are treated as Newtonian point masses.  In order to
test the influence of the binary's mass, we use a range of binary mass
ratios. To simplify the problem we study a binary with mass ratio of
$N$:10~$\msun$ and a 10$~\msun$ interloper, designated as
$N$:$10$:$10$, and vary $N$ between 10$~\msun$ and $10^{3}~\msun$.

The simulations were done using a binary-single scattering code that
was written to be as general purpose as possible.  Because of the vast
parameter space that needs to be covered, the code uses a Monte Carlo
initial condition generator.  The orbits are integrated using {\tt
  hnbody}, a hierarchical, direct N-body integrator, with the adaptive
fourth order Runge Kutta integrator option (K.~Rauch \& D.~Hamilton,
in preparation)\footnote{See http://janus.astro.umd.edu/HNBody/.}.
Because we focus on close approaches where a wide range of timescales
are important, an adaptive scheme is often better than symplectic
methods.

In wide hierarchical triples, direct integration can consume a large
amount of computational time.  To reduce this, we employ a two-body
approximation scheme that tracks the phase of the inner binary.  For a
sufficiently large outer orbit, the orbit is approximately that of an
object about the center of mass of the binary.  We calculate this
approximate two-body orbit analytically and keep track of the inner
binary's phase.  When the outer object nears the binary again, we
revert to direct numerical integration.

The orbit is integrated until one of three conditions is met: 1) one
mass departs along a hyperbolic path, 2) the system forms a
hierarchical triple with outer semimajor axis greater than $2000
\mathrm{~AU}$, an orbit so large that it would likely be perturbed in
the high density of a cluster core and not return, or 3) the
integration is prohibitively long, in which case the encounter is
discarded and restarted with new randomly generated initial
conditions.  Roughly $10^{-4}$ of all encounters had to be restarted
with most occurring for higher mass ratios where resonant encounters
(encounters that have more than one close approach and are not simple
fly-bys) are more common.  In half of our simulations, we evolve the
binary's orbit due to gravitational wave emission after each
encounter.  Since a binary in a cluster spends most of its time and
emits most of its gravitational radiation while waiting for an
encounter rather than during an interaction, we only include
gravitational radiation between encounters.  To isolate this effect,
we run simulations both with and without gravitational radiation.  We
include gravitational radiation by utilizing orbit-averaged
expressions for the change in semimajor axis $a$ and eccentricity $e$
with respect to time \citep{p64}:
\begin{equation}
  \frac{da}{dt} = - \frac{64}{5} 
     \frac{G^{3} m_{0} m_{1} \left(m_{0} + m_{1}\right)}{c^{5} a^{3} 
       \left(1 - e^{2}\right)^{7/2}} 
     \left( 1 + \frac{73}{24}e^{2} + \frac{37}{96}e^{4}\right)
  \label{petersa}
\end{equation}
and
\begin{equation}
  \frac{de}{dt} = - \frac{304}{15} 
     \frac{G^{3} m_{0} m_{1} \left(m_{0} + m_{1}\right)}{c^{5} a^{4} 
       \left(1 - e^{2}\right)^{5/2}} \left(e + \frac{121}{304}e^{3}\right),
  \label{peterse}
\end{equation}
where $m_{0}$ and $m_{1}$ ($m_{0} \ge m_{1}$) are the gravitational
masses of the binary pair.  Here $G$ is the gravitational constant,
and $c$ is the speed of light.  The orbital elements are evolved until
the next encounter takes place, at a time that we choose randomly from
an exponential distribution with a mean encounter time,
$\left<\tau_{\mathrm{enc}}\right> = 1/\left<nv_{\infty}\sigma\right>$,
where $n$ is the number density of objects in the cluster's core,
$v_{\infty}$ is the relative velocity, and $\sigma$ is the
cross-section of the binary.  If we assume the mass of the binary
$m_{0} + m_{1} \gg m_{2}$, then
\begin{equation}
  \sigma \approx \pi r_{p}^{2} + 4\pi r_{p} G
  \left(m_{0} + m_{1}\right) / v_{\infty}^{2},
\label{crosssection}
\end{equation}
where $r_{p}$ is the maximum considered close approach of $m_{2}$ to
the binary's center of mass.  For a thermal distribution of stellar
speeds, $v_{\infty} = \left( m_{\mathrm{avg}} / m_{2} \right)^{1/2}
v_{\mathrm{ms}}$, where $m_{\mathrm{avg}} = 0.4~\msun$ is the average
mass of the main sequence star and $v_{\mathrm{ms}}$ is the main
sequence velocity dispersion.  In our simulations, the second term of
Eq.~\ref{crosssection}, gravitational focusing, dominates over the
first.  Averaging over velocity (assumed to be Maxwellian) we find

\begin{equation}
  \left<\tau_{\mathrm{enc}}\right> = 2 \times 10^{7} 
    \left(\frac{v_{\mathrm{ms}}}{10~\mathrm{km\;s^{-1}}}\right) 
    \left(\frac{10^{6}~\mathrm{pc^{-3}}}{n}\right) 
    \left(\frac{1~\mathrm{AU}}{r_{p}}\right) 
    \left(\frac{1~\msun}{m_{0} + m_{1}}\right)
    \left(\frac{1~\msun}{m_{2}}\right)^{1/2} \mathrm{yr}.
  \label{enctime}
\end{equation}
We then subject the binary to another encounter using orbital
parameters adjusted by both the previous encounter and the
gravitational radiation emitted between the encounters.  This sequence
of encounters continues until the binary merges due to gravitational
wave emission.  If orbital decay is not being calculated, then we
determine that the binary has merged when the randomly drawn encounter
time is longer than the timescale to merger, which is approximately
\begin{equation}
  \tau_{\mathrm{merge}} \approx 6\times10^{17} 
  \frac {\left(1~\msun\right)^{3}} {m_{0} m_{1} \left(m_{0} + m_{1}\right)} 
  \left(\frac{a}{1~\mathrm{AU}}\right)^{4} 
  \left(1 - e^{2}\right)^{7/2} \mathrm{yr}
\label{mergetime}
\end{equation}
for the high eccentricities of importance in this paper.

Global energy and angular momentum are monitored to ensure accurate
integration.  The code also keeps track of the duration of encounters,
the time between encounters, changes in semimajor axis and
eccentricity, and exchanges (events in which the interloping mass
replaces one of the original members of the binary and the replaced
member escapes).

As a test of our code, we compared simulations of several individual
three-body encounters to compare with the work of \citet{hhm96}.  As
part of a series of works examining binary-single star scattering
events, \citet{hhm96} performed numerical simulations of very hard
binaries with a wide range of mass ratios and calculated their
cross-sections for exchange.  We ran simulations of one encounter each
of a sample of mass ratios for comparison.  To facilitate comparison
of encounters with differing masses, semimajor axes, and relative
velocities of hard binaries, \citet{hhm96} use a dimensionless
cross-section,
\begin{equation}
  \bar{\sigma} = \frac {2 v_{\infty}^{2} \Sigma} 
  {\pi G \left(m_{0} + m_{1} + m_{2}\right) a} ,
  \label{sigmabar}
\end{equation}
where $v_{\infty}$ is the relative velocity of the interloper and the
binary's center of mass at infinity and $\Sigma$ is the physical
cross-section for exchanges. We calculate $\Sigma$ as the product of
the fraction of encounters that result in an exchange
($f_{\mathrm{ex}}$) and the total cross-section of encounters
considered: $f_{\mathrm{ex}}\pi b_{\mathrm{max}}^{2}$, where
$b_{\mathrm{max}}$ is an impact parameter large enough to encompass
all exchange reactions.  Our cross-sections are in agreement with
those of \citet{hhm96} within the combined statistical uncertainty as
seen in Table~\ref{exchange}.

\begin{deluxetable}{rccc}
  \footnotesize
  \tablecaption{Single Encounter Cross-sections for Exchange}
 \tablehead{
    \colhead{$m_{0}$:$m_{1}$:$m_{2}$} &
    \colhead{Ejected Mass} &
    \colhead{HHM96} &
    \colhead{This Work} 
  }
  \startdata
10~:~~1~:~~1  &  1 & 1.054 $\pm$ .105& 1.086 $\pm$ .023\\
              & 10 & ----- & ----- \\
10~:~~1~:~10  &  1 & 7.825 $\pm$ .360& 7.741 $\pm$ .255\\
              & 10 & 0.520 $\pm$ .087& 0.513 $\pm$ .043\\
~3~:~~1~:~~1  &  1 & 2.311 $\pm$ .170& 2.465 $\pm$ .073\\
              &  3 & 0.059 $\pm$ .025& 0.072 $\pm$ .007\\

  \enddata
  \label{exchange}
  \tablecomments{ This table compares dimensionless cross-sections for
    exchange $\bar{\sigma}$ (see text for details) calculated by
    \citealt{hhm96} and by us.  The first column lists the masses,
    with binary components $m_{0}$ and $m_{1}$.  Column two shows the
    mass of the ejected object.  The ejection of the smaller mass is
    energetically favored so it always has a larger cross-section.
    There is general agreement between the two calculations to within
    the statistical uncertainty, which we calculate as
    $\bar{\sigma}/N_{\mathrm{ex}}^{1/2}$, where $N_{\mathrm{ex}}$ is
    the total number of exchanges.}

\end{deluxetable}

\section{Simulations and Results}
\label{simulations}

We used our code to run numerical experiments of three-body encounter
sequences with a variety of mass ratios.  The binaries consisted of a
dominant body with mass, $m_{0} =$ 10, 20, 30, 50, 100, 200, 300, 500,
or 1000~$\msun$ and a secondary of mass $m_{1} = 10~\msun$.  Because
of mass segregation, the objects that the binary encounters will be
the heaviest objects in the cluster.  In order to simplify the
problem, we consider only interactions with interlopers of mass $m_{2}
= 10~\msun$.  The binary starts with a circular $a=10~\mathrm{AU}$
orbit, and the interloper has a relative speed at infinity of
$v_{\infty} = 10~\mathrm{~km~s^{-1}}$ and an impact parameter, $b$,
relative to the center of mass of the binary such that the pericenter
distance of the hyperbolic encounter would range from $r_{p} = 0$ to
$5a$.  For all binaries, $v_{\mathrm{circ}} = \left[G\left(m_{0} +
    m_{1}\right)/a\right]^{1/2} \ge 40\kms \gg v_{\infty}$, and thus
all are considered hard.  The Monte Carlo initial condition generator
distributes the orientations and directions of encounters
isotropically in space, and the initial phase of the binary is
randomized such that it is distributed equally in time.  We assume the
cluster core has a density of $n = 10^{5}~\mathrm{pc}^{-3}$ and an
escape velocity of $v_{\mathrm{esc}}=50\kms$ for the duration of the
simulation.  We discuss the consequences of changing the escape
velocity in \S~\ref{imbhformation}.  For each mass ratio, we simulate
1000 sequences with and without gravitational radiation between
encounters.

\subsection{Pure Newtonian Sequences}
\label{newt}

\begin{figure}
\epsscale{0.9}
\plottwo{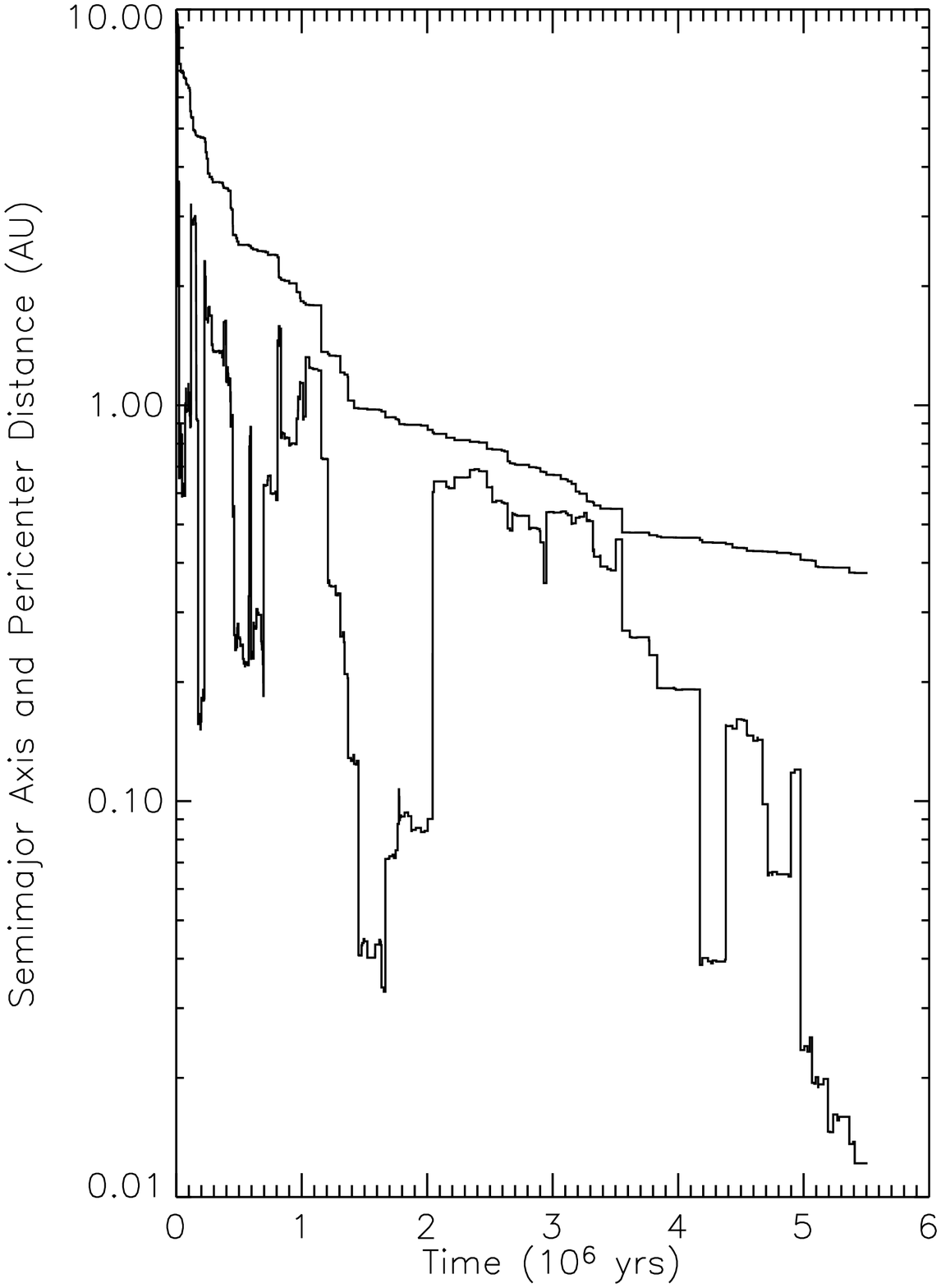}{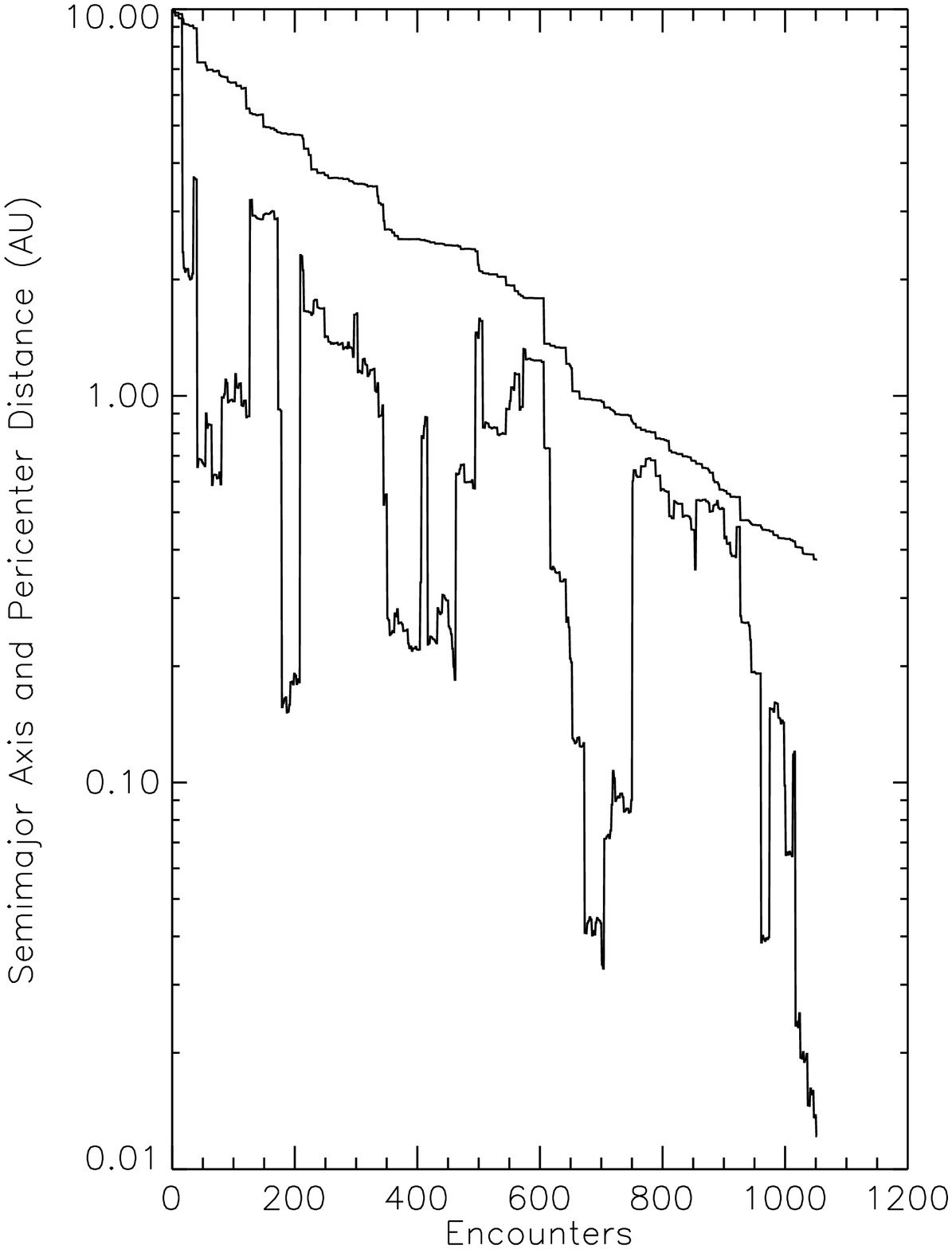}

\caption{Newtonian 1000:10:10 sequence.  These panels show the
  semimajor axes (upper lines) and pericenter distances (lower lines)
  as functions of time (left panel) and number of encounters (right
  panel) for one sequence of encounters with no gravitational wave
  emission.  Each change in $a$ and $r_{p}$ is the result of a
  three-body encounter.  Since the binary is hard, the semimajor axis
  gradually tightens by a roughly constant fractional amount per
  encounter with most of the time spent hardening the final fraction
  when close encounters are rare.  The pericenter distance, however,
  fluctuates greatly due to large changes in eccentricity during a
  single encounter.  The sequence ends at a very high eccentricity
  when the binary would merge due to gravitational radiation before
  the next encounter.}

\label{newtlife}
\end{figure}

Figure~\ref{newtlife}a shows the change of semimajor axis and
pericenter distance as a function of time over the course of a typical
Newtonian sequence.  The encounters themselves take much less time
then the period between encounters, so a binary spends virtually all
its time waiting for an interloper.  Most of the time in this example
is spent hardening the orbit from $1\au$ to $0.4\au$ because as the
binary shrinks, its cross-section decreases and the timescale to the
next encounter increases.  Figure~\ref{newtlife}b shows the same
sequence plotted as a function of number of encounters.  The semimajor
axis decreases by a roughly constant factor with each encounter.  This
is expected for a hard binary, which, according to Heggie's Law
\citep{h75}, tends to harden with each encounter at a rate independent
of its hardness.  The eccentricity and therefore the pericenter
distance, $r_{p} = a\left(1-e\right)$, however, can change
dramatically in a single encounter \citep[for a discussion on
eccentricity change of a binary in a cluster, see][]{hr96}.  This
sequence ends with a very high eccentricity ($e=0.968$), which reduces
the merger time given by Eq.~\ref{mergetime} to less than
$\tau_{\mathrm{enc}}$.

Table~\ref{coreresults} summarizes our main results and shows a number
of interesting trends.  The average number of encounters per sequence,
$\left<n_{\mathrm{enc}}\right>$, increases with increasing mass ratio
since the energy that the interloper can carry away scales as $\Delta
E / E\sim m_{1}/\left(m_{0} + m_{1}\right)$ \citep{q96} and since
$n_{\mathrm{enc}} \sim E / \Delta E$ for a constant eccentricity.
Energy conservation assures that every hardening event results in an
increased relative velocity between the binary and the single black
hole.  If the velocity of the single black hole relative to the
barycenter, and thus the globular cluster, is greater than the escape
velocity of the cluster core \citep[typically $v_{\mathrm{esc}} = 50
\kms$ for a dense cluster; see][]{w85}, then the single mass will be
ejected from the cluster.  The average number of ejected masses per
sequence, $\left<n_{\mathrm{ej}}\right>$, also increases with
increasing mass ratio because the higher mass ratio sequences have a
larger number of encounters and because the larger mass at a given
semimajor axis has more energy for the interloper to tap.
Conservation of momentum guarantees that when a mass is ejected from
the cluster at very high velocity, the binary may also be ejected.
Table~\ref{coreresults} lists $\left<f_{\mathrm{binej}}\right>$, the
fraction of sequences that result in the ejection of the binary from
the cluster.  As expected, the fraction decreases sharply with
increasing mass such that virtually none of the binaries with mass
greater than $300~\msun$ escape the cluster.

\begin{deluxetable}{rlrrrrrr}
  \footnotesize
  \tablecaption{Sequence Statistics }
  \tablehead{
    \colhead{$m_{0}$} &
    \colhead{Case} &
    \colhead{$\left<n_{\mathrm{enc}}\right>$} &
    \colhead{$\left<n_{\mathrm{ej}}\right>$} &
    \colhead{$\left<f_{\mathrm{binej}}\right>$} &
    \colhead{$\left<t_{\mathrm{seq}}\right>/10^{6}~\mathrm{yr}$} &
    \colhead{$\left<a_{f}\right>/\mathrm{AU}$} &
    \colhead{$\left<e_{f}\right>$} 
  }
  \startdata
  10 & Newt.    &  51.6 &   3.9 & 0.880 & 82.72  &      0.164 &      0.929 \\
     & {\bf GR Evol.} &  {\bf 48.7} &   {\bf 3.7} & {\bf 0.839} & {\bf 59.89}  &      {\bf 0.190} &      {\bf 0.901} \\
  20 & Newt.    &  51.3 &   6.5 & 0.835 & 65.94  &      0.178 &      0.924 \\
      & {\bf  GR Evol.} & {\bf 47.1} & {\bf 6.1} & {\bf 0.776} & {\bf 43.46} & {\bf 0.230} & {\bf 0.898} \\
  30  &  Newt.    &  58.9 &   9.3 & 0.753 & 49.11  &      0.198 &      0.926 \\
      & {\bf GR Evol.} & {\bf 55.1} & {\bf 8.6} & {\bf 0.676} & {\bf 31.89} & {\bf 0.222} & {\bf 0.892} \\
  50 & Newt.    &  73.2 &  14.6 & 0.581 & 33.75  &      0.230 &      0.919 \\
      & {\bf GR Evol.} & {\bf 66.7} & {\bf 13.0} & {\bf 0.455} & {\bf 22.73} & {\bf 0.285} & {\bf 0.892} \\
 100 & Newt.    & 102.0 &  24.0 & 0.229 & 21.35  &      0.327 &      0.936 \\
      & {\bf  GR Evol.} & {\bf   93.4} & {\bf   20.1} & {\bf  0.161} & {\bf  14.97} & {\bf       0.357} & {\bf       0.873} \\
 200 & Newt.    & 158.4 &  38.2 & 0.043 & 15.13  &      0.387 &      0.938 \\
      & {\bf  GR Evol.} & {\bf  140.3} & {\bf   31.5} & {\bf  0.026} & {\bf   9.998} & {\bf       0.444} & {\bf       0.872} \\
 300 & Newt.    & 208.5 &  49.1 & 0.013 & 11.89  &      0.468 &      0.943 \\
      & {\bf  GR Evol.} & {\bf  184.0} & {\bf   39.4} & {\bf  0.006} & {\bf   7.822} & {\bf       0.445} & {\bf       0.874} \\
 500 & Newt.    & 308.7 &  71.1 & 0.001 &  9.920 &      0.528 &      0.944 \\
      & {\bf  GR Evol.} & {\bf  269.1} & {\bf   54.9} & {\bf      0} & {\bf   6.225} & {\bf       0.488} & {\bf       0.860} \\
1000 & Newt.    & 562.4 & 117.3 &     0 &  7.363 &      0.641 &      0.953 \\
      & {\bf  GR Evol.} & {\bf  483.0} & {\bf   88.9} & {\bf      0} & {\bf   4.427} & {\bf       0.556} & {\bf       0.851}
  \enddata
  \label{coreresults}
  \tablecomments{ Table~\ref{coreresults} summarizes the main results
    of our simulations of sequences of three-body encounters.  For
    each dominant mass, $m_{0}$, we ran 1000 sequences of pure
    Newtonian encounters (Newt.) and 1000 sequences of the more
    realistic Newtonian encounters with gravitational radiation
    between encounters (GR Evol.).  The columns list the average
    number of encounters per sequence $\left<n_{\mathrm{enc}}\right>$,
    the average number of black holes ejected from the cluster in each
    sequence $\left<n_{\mathrm{ej}}\right>$, the fraction of sequences
    in which the binary is ejected from the cluster,
    $\left<f_{\mathrm{binej}}\right>$, the average total time for the
    sequence $\left<t_{\mathrm{seq}}\right>$, the average final
    semimajor axis $\left<a_{f}\right>$, and the average final
    eccentricity $\left<e_{f}\right>$.}
\end{deluxetable}

\begin{figure}
\epsscale{0.75}
\rotatebox{90}{\plotone{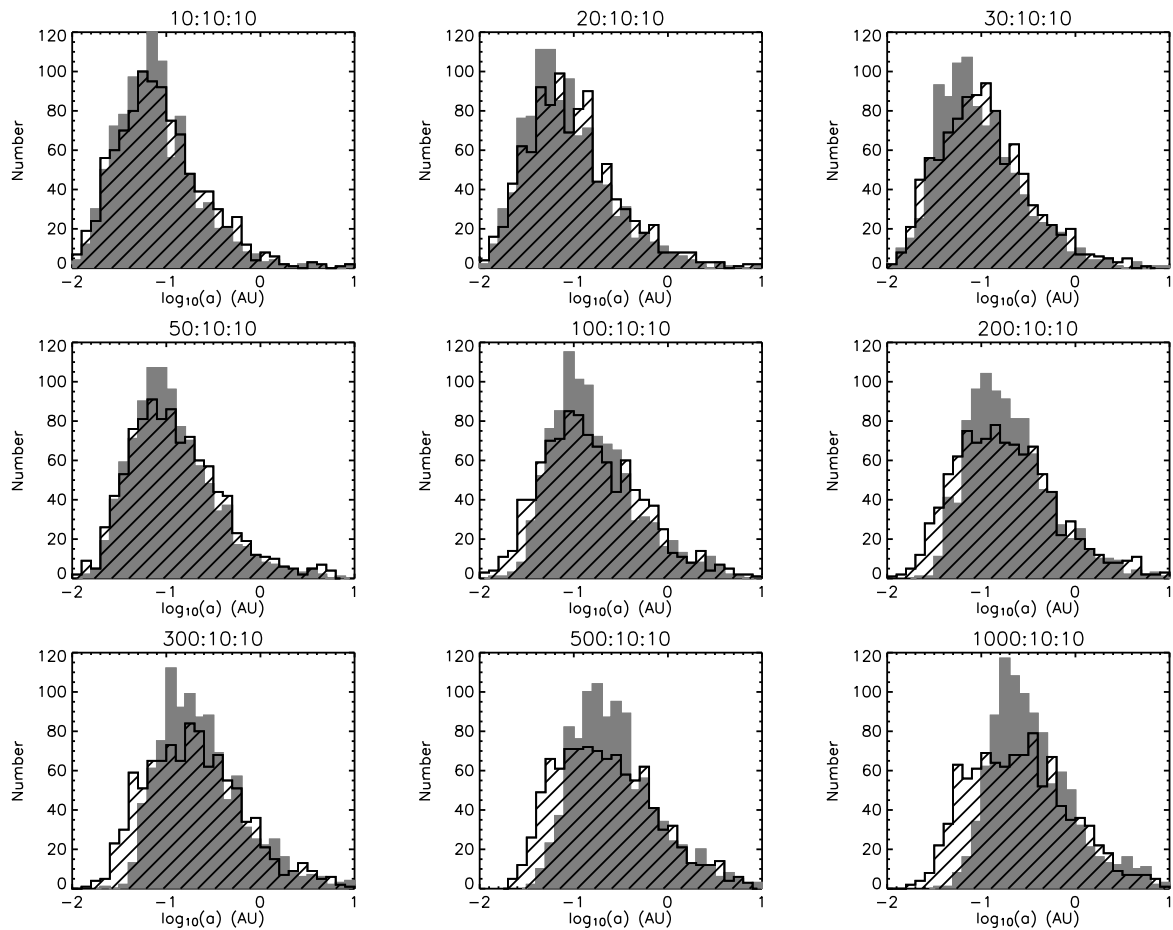}}
\caption{Histograms of final semimajor axes for all mass ratios.  The
  solid histograms are pure Newtonian sequences, and the hatched
  histograms are sequences with gravitational radiation between
  encounters.  The histograms all have similar shapes with a sharp
  drop at low $a$ since the binary tends to merge before another
  encounter can harden it, and it has long tail at high $a$ where the
  binary will only merge with high eccentricity.  The sequences with
  gravitational radiation have falloffs at smaller $a$ than those
  without due to both the circularization and the extra source of
  hardening.}
\label{ahisto}
\end{figure}

The shape and size of the orbit after its last encounter determine the
dominant gravitational wave emission during the inspiral and are of
particular interest to us.  The distribution of pre-merger semimajor
axes for all mass ratios is shown in Figure~\ref{ahisto}.  The
distributions all have a similar shape that drops off at low $a$
because the binary tends to merge before another encounter can harden
it.  For large orbits the binary will only merge for a high
eccentricity, and thus there is a long tail in the histograms towards
high $a$ from encounters that resulted in an extremely high
eccentricity.  The distributions for lower mass ratios are shifted to
smaller $a$ because for a given orbit, a less massive binary will take
longer to merge.  This can also be seen in the mean final semimajor
axis, $\left<a_{f}\right>$, in Table~\ref{coreresults}.

\begin{figure}
\epsscale{0.75}
\rotatebox{90}{\plotone{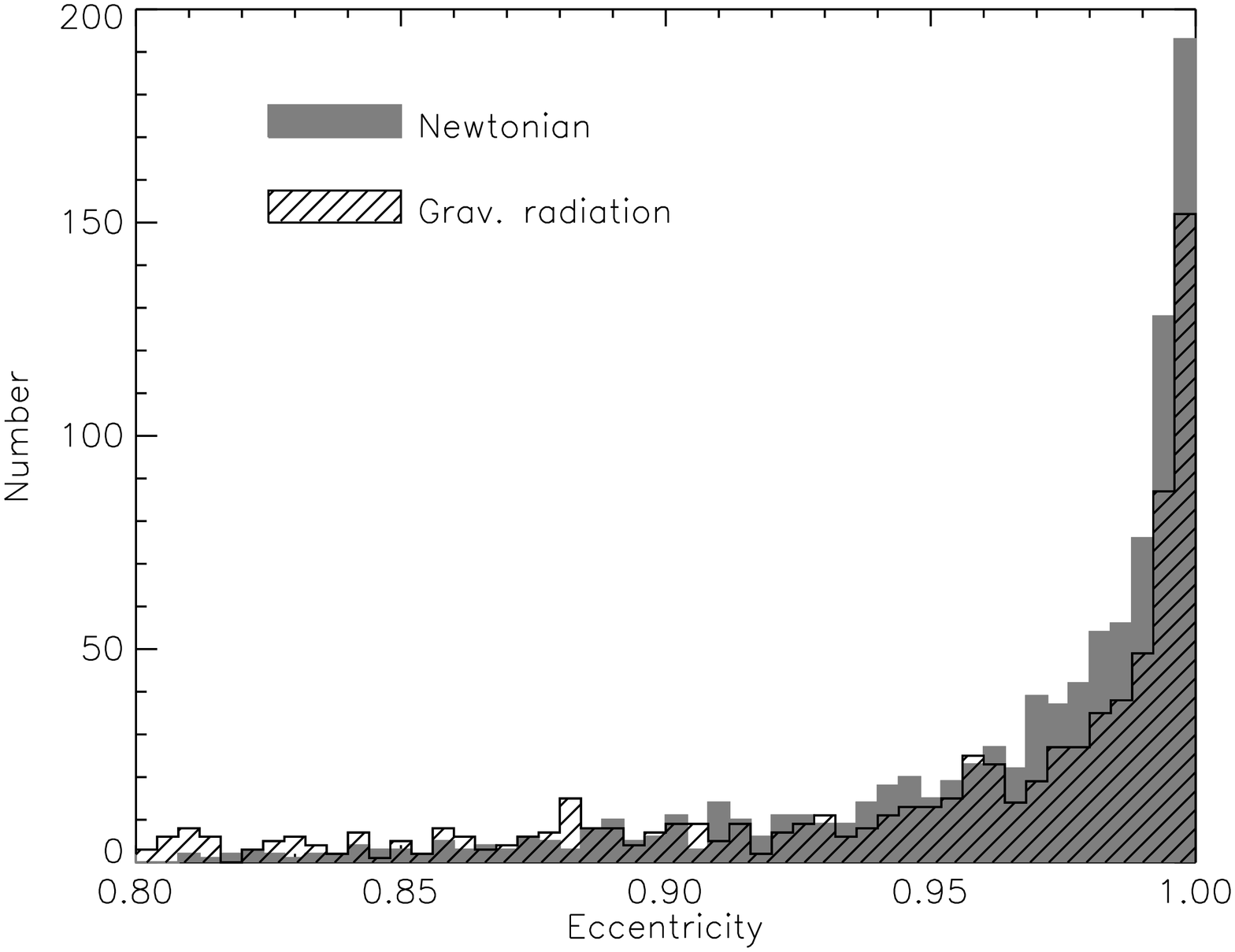}}
\caption{Histogram of final eccentricities for 1000:10:10 mass ratio.
  The solid histogram is from pure Newtonian sequences, and the
  hatched histogram is from sequences with gravitational radiation
  between encounters.  The histogram is cut at $e = 0.8$ because
  $e_{f} < 0.8$ is rare.  The histograms have roughly the same shape
  for both cases and for all mass ratios although the gravitational
  wave sequences have a consistently lower mean at higher mass ratios
  because gravitational wave emission damps eccentricities.  The
  histograms show a decidedly non-thermal distribution and are
  strongly peaked near $e=1$.  Because the timescale to merge due to
  gravitational radiation is so strongly dependent on $e$, the binary
  will merge when it happens to reach a high eccentricity.}
\label{ehisto}
\end{figure}

\begin{figure}
\epsscale{0.75}
\rotatebox{90}{\plotone{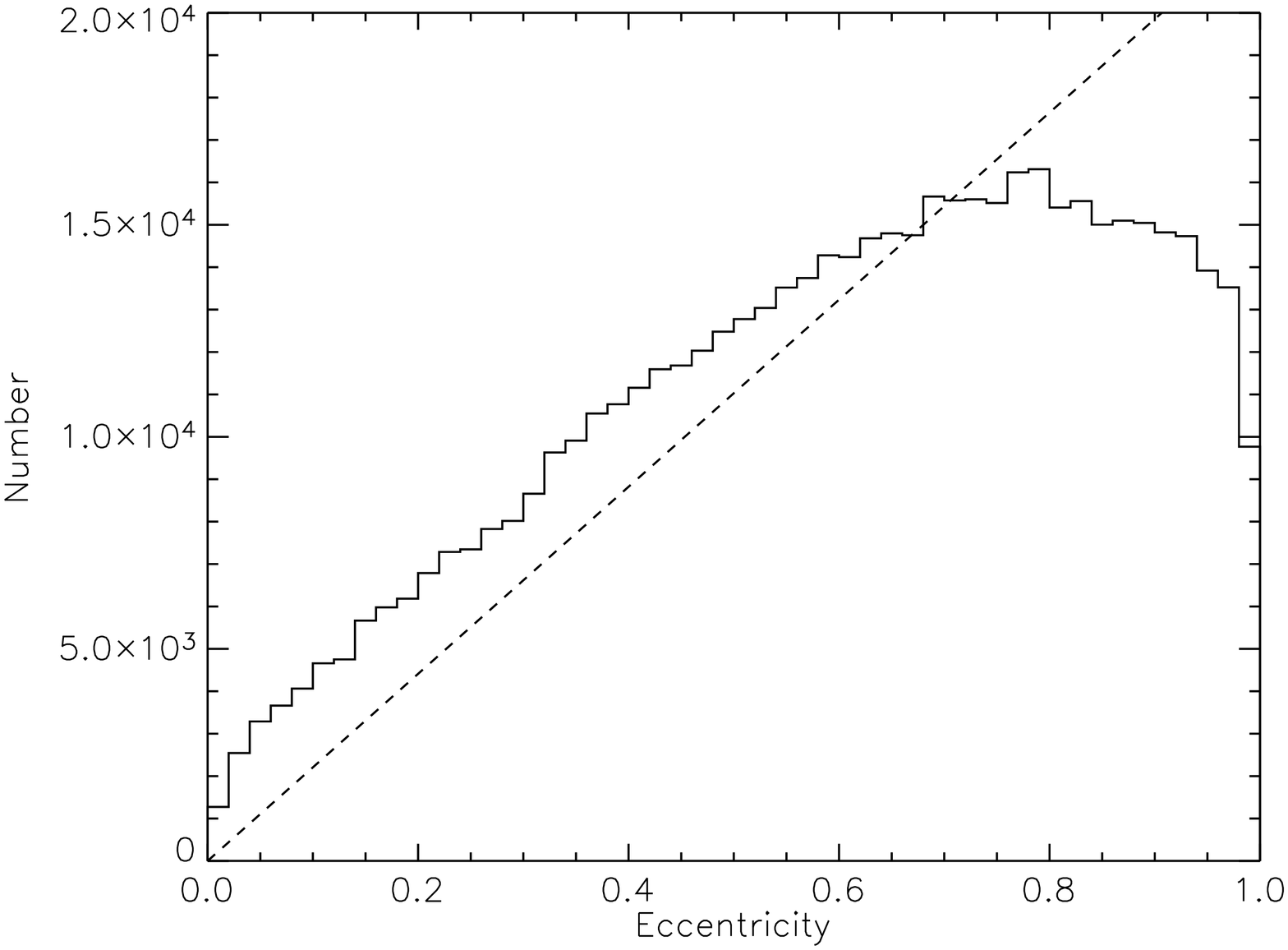}}
\caption{ Solid line is a histogram of all eccentricities after each
  encounter except for the first ten for all pure Newtonian sequences
  of 1000:10:10.  The dashed line is a thermal distribution of
  eccentricities.  The distribution is roughly thermal for low
  eccentricity but deviates for $e \ga 0.6$.  The expected
  thermal distribution of eccentricities is altered by losses of high
  eccentricity orbits to merger.}
\label{thermale}
\end{figure}

Figure~\ref{ehisto} shows the distribution of binary eccentricities
after the final encounter for one mass ratio.  The plot is strongly
peaked near $e=1$, a property shared by all other mass ratios.  This
distribution is definitely not thermal, which would have a mean
eccentricity $\left<e\right>_{\mathrm{th}} \approx 0.7$.  The high eccentricity
before merger results from both the strong dependence of merger time
on eccentricity and the fact that the eccentricity can change
drastically in a single encounter (see Fig.~\ref{newtlife}).  As the
semimajor axis decreases by roughly the same fractional amount in each
encounter, the eccentricity increases and decreases by potentially
large amounts with each strong encounter.  When the eccentricity
happens to reach a large value, the binary will merge before the next
encounter.  Figure~\ref{thermale} shows the eccentricity distribution
for all encounters after the first 10 for all 1000 sequences with a
mass ratio of 1000:10:10.  The distribution is roughly thermal up to
high eccentricity where the binaries merge.  Thus merger selectively
removes high eccentricity binaries from a thermal distribution.

\subsection{General Relativistic Binary Evolution}
\label{grseq}
\begin{figure}
\epsscale{0.75} \rotatebox{90}{\plotone{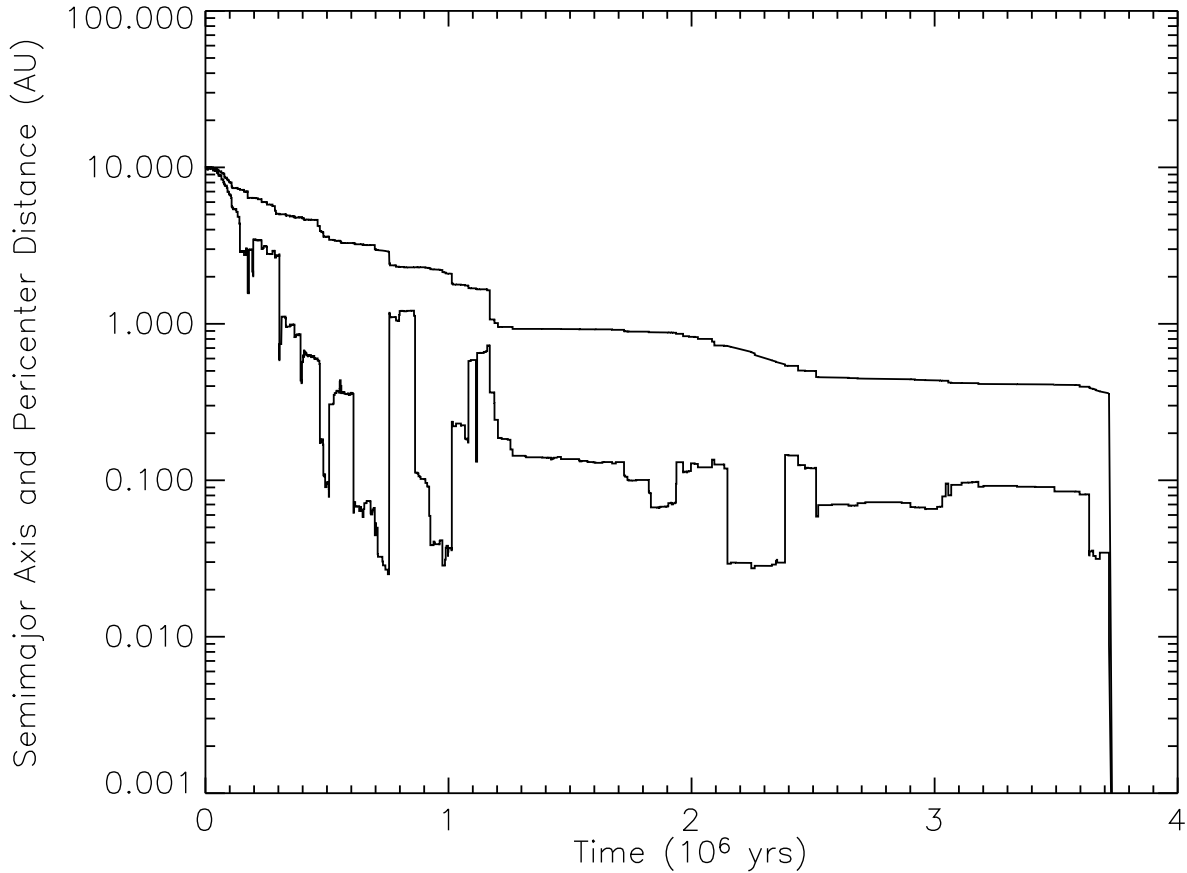}}
\caption{1000:10:10 gravitational radiation sequence.  
  Same as Figure~\ref{newtlife}a but for a sequence with gravitational
  radiation between encounters.  The effects of gravitational
  radiation can be seen between $2.2$ and $2.4\times 10^{6}$~years.
  Over this period, the binary undergoes about ten interactions that
  do not significantly affect its orbit.  During this time, the
  semimajor axis decays from $a=0.713\au$ to $0.550\au$ while the
  pericenter distance remains small and roughly constant.  When an
  encounter reduces the eccentricity at $2.4\times 10^{6}$ years,
  gravitational radiation is strongly reduced.  Gravitational
  radiation becomes important again at the end of the sequence.  The
  sequence ends with the binary's merger from gravitational waves.}
\label{grlife}
\end{figure}

The addition of gravitational radiation between Newtonian encounters
is expected to alter a sequence since it is an extra source of
hardening and since it circularizes the binary.  Figure~\ref{grlife}
shows a typical sequence for the 1000:10:10 mass ratio including
gravitational radiation.  Three-body interactions drive the binary's
eccentricity up to $e=0.959$ and its semimajor axis down to
$a=0.713\au$.  Then starting at $t = 2.2\times 10^{6}~\mathrm{yr}$
over the course of about ten interactions that only weakly affect the
eccentricity and semimajor axis, gravitational radiation causes the
orbit to decay to $a=0.550\au$ and $e=0.946$ while the pericenter
distance remains roughly constant.  The corresponding semimajor axis
change in the Newtonian only sequences in Figure~\ref{newtlife} takes
45 encounters and more than twice as long although one must be careful
when comparing two individual sequences.  Gravitational waves make the
most difference when the pericenter distance is small, which is
guaranteed at the end of a sequence, but can also happen in the middle
as Figure~\ref{grlife} shows.

Table~\ref{coreresults} summarizes the effect of adding gravitational
radiation.  In general the effect is greater at higher masses because
gravitational radiation is stronger for a given orbit.  Because of the
extra energy sink, the binaries merge with fewer encounters, fewer
black holes are ejected, and the fraction of sequences in which a
binary is ejected is smaller.  The most dramatic change is in the
duration of the sequence, which gravitational radiation reduces by
27\% to 40\%.  The distributions of final semimajor axes
(Fig.~\ref{ahisto}) and final eccentricities (Fig.~\ref{ehisto}) have
similar shapes to the Newtonian only distributions.  Due to the
circularizing effect of gravitational radiation, binaries of all mass
ratios merge with a smaller $\left<e_{f}\right>$ than Newtonian only
sequences with the largest difference at high mass ratios.
Gravitational radiation also produces a smaller $\left<a_{f}\right>$
for $m_{0} \ga 300~\msun$.  This can be seen in Figure~\ref{ahisto}
where the gravitational radiation simulations display an excess number
of sequences with low $a_{f}$, which is a consequence of the binaries'
lower $e_{f}$.

\section{Implications for IMBH Formation}
\label{imbhformation}

We can use these simulations to test the \citet{mh02} model of IMBH
formation.  We assume that a $50~\msun$ seed black hole with a
$10~\msun$ companion will undergo repeated three-body encounters with
$10~\msun$ interloping black holes in a globular cluster with
$v_{\mathrm{esc}}=50\kms$ and $n=10^{5}~\mathrm{pc}^{-3}$.  We also
assume that the density of the cluster core remains constant as the
IMBH grows.  We then test whether the model of \citet{mh02} can build
up to IMBH masses, which we take to be $10^{3}~\msun$, 1) without
ejecting too many black holes from the cluster, 2) without ejecting
the IMBH from the cluster, and 3) within the lifetime of the globular
cluster.  We also test how these depend on escape velocity and seed
mass.

\begin{figure}
\epsscale{0.75}
\rotatebox{90}{\plotone{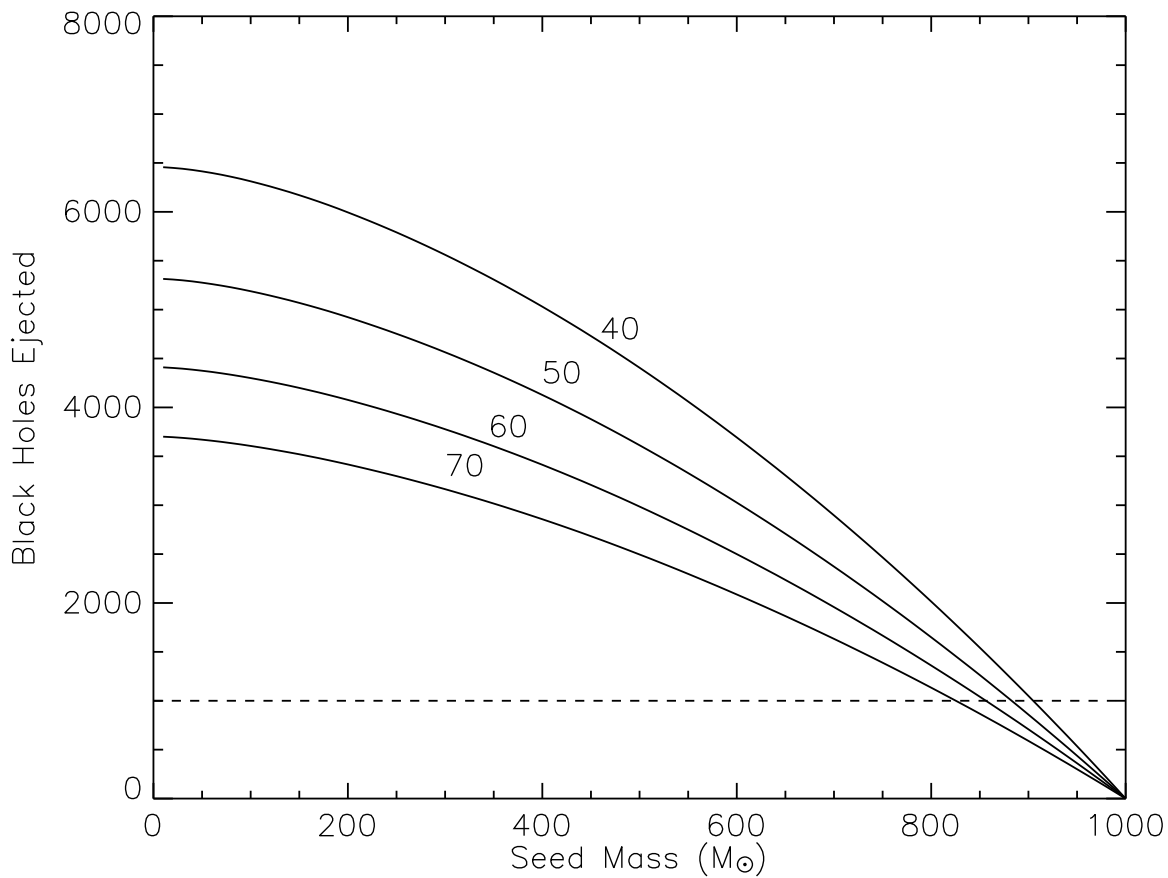}}
\caption{Plot of total number of black holes ejected in building up to
  $1000~\msun$ as a function of seed mass for the gravitational
  radiation case assuming different escape velocities.  The four
  curves show different assumed cluster escape velocities in\kms.  For
  all but the largest seed masses, the number of black holes ejected
  is greater than the estimated $\sim 10^{3}$ (indicated by the dashed
  line) present in a young globular cluster.  }

\label{ejvm}
\end{figure}

If the number of black holes ejected is greater than the total number
of black holes in the cluster core, then the IMBH cannot build up to
the required mass by accreting black holes alone.  To calculate the
total number of black holes ejected while building up to large masses,
we sum the average number of ejections using a linear interpolation of
the values in Table~\ref{coreresults}.  Assuming a cluster escape
velocity of $v_{\mathrm{esc}} = 50\kms$, we find that the total number
of black holes ejected when building up to $1000~\msun$ is
approximately 6800 for our Newtonian only and 5300 for gravitational
radiation simulations.  This is far greater than the estimated
$10^{2}$ to $10^{3}$ black holes available \citep{pzm00}.  If there
were initially one thousand $10~\msun$ black holes in the cluster,
mergers of the massive black hole with a series of $10~\msun$ black
holes would exhaust half of the black holes in $\sim 2.6 \times
10^{8}~\mathrm{yr}$ and would ultimately produce a $240~\msun$ black
hole.  Increasing the seed mass increases the final mass of the IMBH
when half of the field black holes run out.  If the seed mass were
$100$, $200$, or $300~\msun$, then the model would produce a $270$,
$330$, or $410~\msun$ black hole after exhausting half of the cluster
black hole population in $1.9$, $1.1$, or
$0.8\times10^{8}~\mathrm{yr}$, respectively.  Figure~\ref{ejvm} shows
the number of black holes ejected as a function of initial black hole
mass for a range of escape velocities.  Growth times are much shorter
than the $\sim 10^{9}~\mathrm{yr}$ necessary for stellar-mass black
holes to eject each other from the cluster (\citealt{sh93, pzm00};
J.~M. Fregeau, S.~A. Rappaport, \& V. Corless, in preparation;
R.~O'leary et al., in preparation).  Therefore, self-depletion of
stellar-mass black holes is not a limiting factor.

\begin{figure}
\epsscale{0.75}
\rotatebox{90}{\plotone{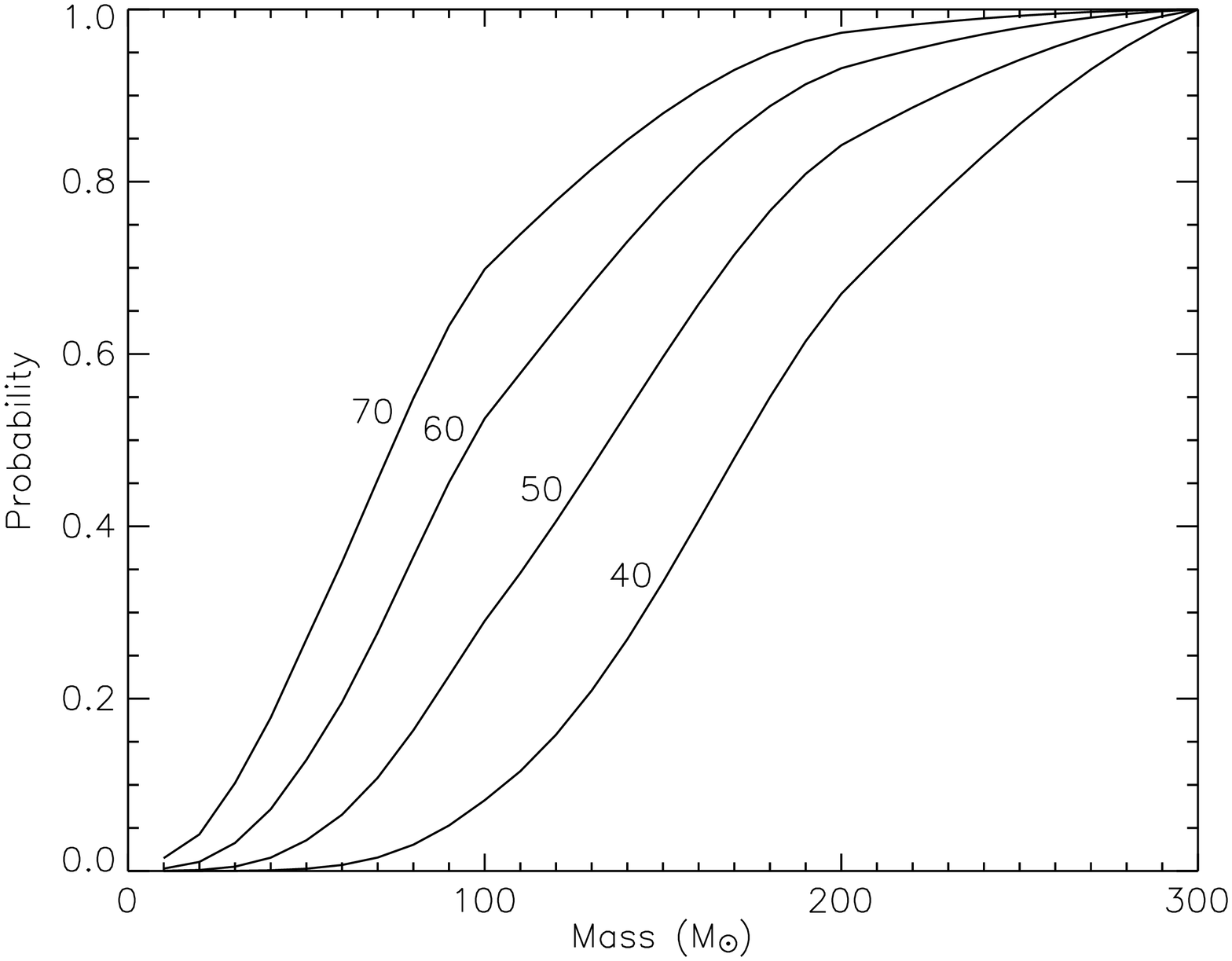}}
\caption{ Plot of an IMBH's probability of remaining in the cluster
  and building up to $300~\msun$ as a function of starting mass of the
  dominant black hole for the gravitational radiation case assuming
  different escape velocities labeled in\kms.  Once the black hole has
  built up to $300~\msun$ it is very unlikely that it will be ejected
  from the cluster.  The lowest mass binaries are much more readily
  ejected and thus are very unlikely to survive a sequence of
  encounters.  \citet{mh02} suggest that IMBHs can be built in this
  manner with a starting mass $\approx 50~\msun$.  We find that such
  small initial masses are likely to be ejected from the cluster core
  for reasonable escape velocities of dense clusters.}
\label{cumprob}
\end{figure}

Of particular concern is whether the three-body scattering events will
eject the binary from the cluster.  The black hole can only merge with
other black holes while it is in a dense stellar environment.  The
probability of remaining in the cluster after one sequence is $P = 1 -
\left<f_{\mathrm{binej}}\right>$.  As can be seen in
Table~\ref{coreresults}, once the black hole has built up to $\sim
300~\msun$, it is virtually guaranteed to remain in the cluster.  When
starting with 50~\msun, we calculate the total probability of building
up to $300~\msun$ to be 0.0356.  Figure~\ref{cumprob} shows the
probability of building up to $300~\msun$ as a function of starting
mass for different escape velocities for the gravitational radiation
case.  Table~\ref{imbhtable} lists probabilities for selected seed
masses and escape velocities for the gravitational radiation case.

\begin{deluxetable}{ccccc}
  \footnotesize
  \tablecaption{IMBH Formation}
 \tablehead{
    \colhead{Seed Mass}&
    \colhead{$v_{\mathrm{esc}}$}&
    \colhead{Probability to remain} &
    \colhead{Number of} &
    \colhead{Time} \\
    \colhead{($\msun$)}&
    \colhead{($\mathrm{km}~\mathrm{s}^{-1}$)}&
    \colhead{in cluster} &
    \colhead{BH ejections} &
    \colhead{($10^{8}~\mathrm{yr}$)}
  }
  \startdata
  50.0  &    40.0  &   0.00264 &  6414   &  7.06 \\
        &    50.0  &   0.0356  &  5276   &   \\
        &    60.0  &   0.129   &  4038   &   \\
        &    70.0  &   0.269   &  3573   &   \\
\tableline
 100.0  &    40.0  &   0.0821  &  6312   &  6.15 \\
        &    50.0  &   0.290   &  5188   &   \\
        &    60.0  &   0.525   &  3963   &   \\
        &    70.0  &   0.698   &  3606   &   \\
\tableline
 200.0  &    40.0  &   0.670   &  5995   &  4.93 \\
        &    50.0  &   0.842   &  4922   &   \\
        &    60.0  &   0.932   &  4077   &   \\
        &    70.0  &   0.978   &  3417   &   \\
\tableline
 300.0  &    40.0  &   1.000   &  5561   &  4.05 \\
        &    50.0  &   1.000   &  4564   &   \\
        &    60.0  &   1.000   &  3777   &   \\
        &    70.0  &   1.000   &  3164   &   
  \enddata
  \label{imbhtable}
  \tablecomments{This table lists values for selected seed masses and
    cluster escape velocities for the gravitational radiation case.
    Column 3 lists the probability for the IMBH to remain in the
    cluster until it reaches a mass of $300~\msun$.  The fourth column
    lists the total number of black holes ejected in building up to
    $1000~\msun$.  Column 5 lists the total time to build up to
    $1000~\msun$. The total time is not affected by the escape
    velocity because the density of black holes in the cluster core is
    taken to be constant.}
\end{deluxetable}

\begin{figure}
\epsscale{0.75}
\rotatebox{90}{\plotone{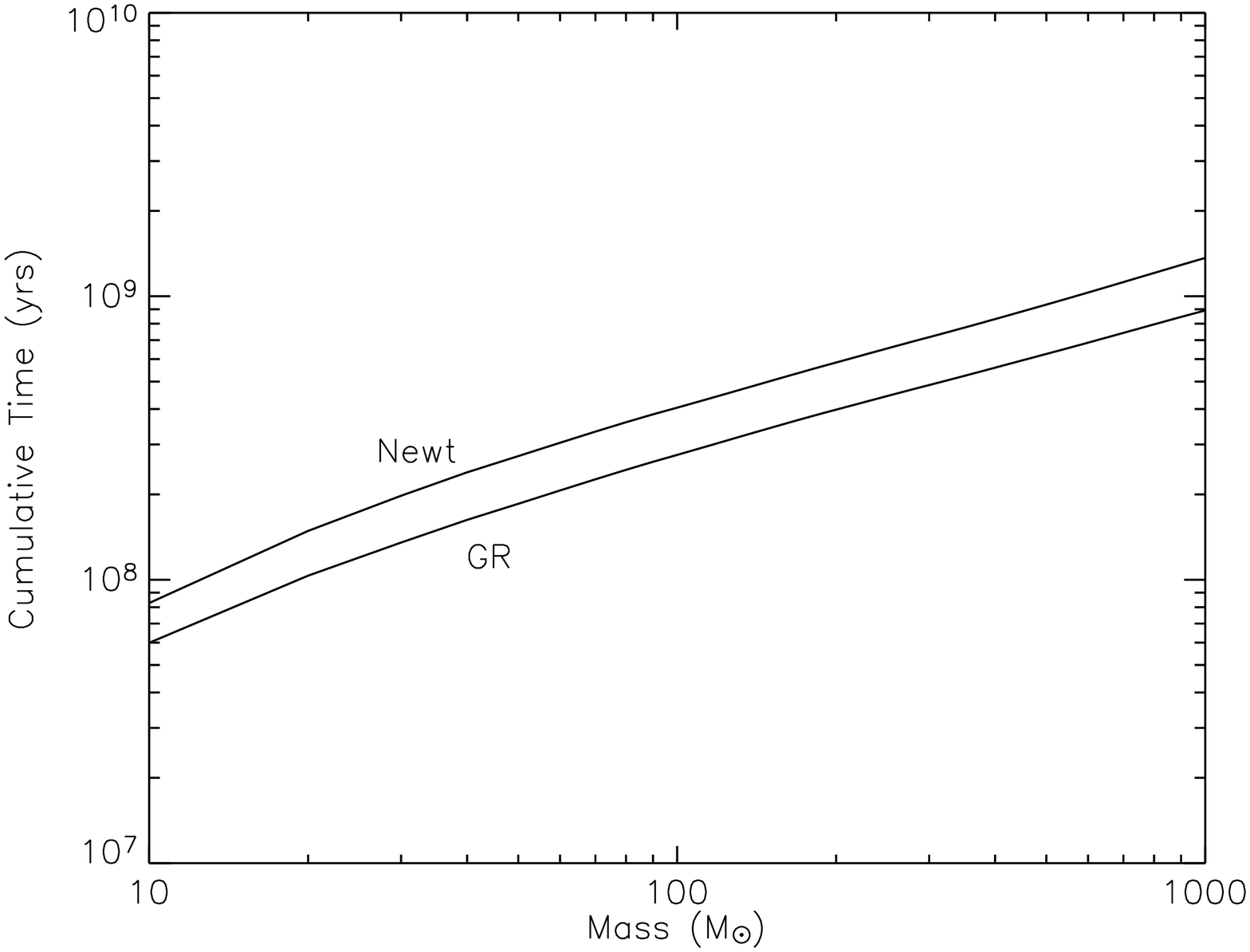}}
\caption{Plot of total time to build up to a certain mass when built
  by mergers with $10~\msun$ black holes for Newtonian only results
  and for runs with gravitational radiation between encounters.  The
  Newtonian only simulations are slower to build up, but both cases
  reach $1000~\msun$ within about $10^{9}$~years.  The time plotted
  assumes a constant density of black holes for the duration of IMBH
  formation. }
\label{timescale}
\end{figure}

In a similar manner, we calculate the total time to build up to
$1000~\msun$, assuming that the supply of stellar-mass black holes and
density remain constant, an assumption which leads to an
underestimation of the time.  While the time per merger is larger for
the smaller masses, the total time is dominated at the higher masses
since more mergers are needed for the same fractional increase in
mass.  For Newtonian only simulations the total time is
$1.1\times10^{9}~\mathrm{yr}$, and for simulations with gravitational
radiation the total time is $7.1\times10^{8}~\mathrm{yr}$.  These are
much less than the age of the host globular clusters.
Figure~\ref{timescale} shows the time to reach a specified mass for
both the Newtonian and gravitational radiation cases.

Although there is clearly enough time to build IMBHs as \citet{mh02}
propose, the issues of whether there are enough stellar-mass black
holes and whether the cluster will hold onto the IMBH remain.  The
combination of an initial mass of $50~\msun$ and an escape velocity of
$50\kms$ is not likely to produce an IMBH in a globular cluster
through three-body interactions with $10~\msun$ black holes, but the
general process could still produce IMBHs.  \citet{mh02} argued that a
seed mass of 50$~\msun$ would be retained, but for analytical
simplicity they assumed that every encounter changed the semimajor
axis by the same fractional amount $\left<\Delta a / a \right>$.  Some
encounters, however, can decrease the semimajor axis by several times
the average value and thus impart much larger kicks.  The authors
therefore underestimated the minimum initial mass necessary to remain
in the cluster.  A hierarchical merging of stellar-mass black holes
could, however, still produce an IMBH if 1) the initial mass of the
black hole were were greater than $50~\msun$, 2) the escape velocity
of the cluster were greater than $50\kms$, or 3) additional dynamics
were involved.  We consider each of these in turn.

If the mass of the initial black hole were, e.g., $250~\msun$ before
the onset of compact object dynamics, dynamical kicks would not be
likely to eject the IMBH, and it would require fewer mergers to reach
$1000~\msun$ and thus a smaller population of stellar mass black
holes.  The initial black hole could start with such a mass if it
evolved from a massive Population~III star or from a runaway collision
of main sequence stars \citep{pzm02, gfr04}, or it could reach such
a mass by accretion of young massive stars, which would be torn apart
by tidal forces and impart little dynamical kick.

If the initial globular cluster mass is high enough (work by
\citealt{m2etal01} indicate masses of $10^{7}~\msun$ are available),
then the cluster's gravity may be strong enough to retain the gas
normally expelled by the first generation of supernovae.  If that
increases the escape velocity to, e.g., $v_{\mathrm{esc}}=70\kms$, the
interactions result in a smaller fraction of ejected binaries.  The
probability of building from $50~\msun$ to $1000~\msun$ then increases
by almost an order of magnitude.

In addition, processes with lower dynamical kicks could prevent
ejection.  One promising mechanism is the Kozai resonance \citep{k62,
mh02b}.  If a stable hierarchical triple is formed, then resonant
processes can pump up the inner binary's eccentricity high enough so
that it would quickly merge due to gravitational radiation and without
any dynamical kick to eject the IMBH from the cluster.  Two-body
captures (captures in which an interloper passes close enough to the
isolated IMBH that it becomes bound and merges due to gravitational
radiation) would also result in mergers without dynamical kicks.  Both
Kozai-resonance-induced mergers and two-body captures are devoid of
dynamical kicks, but they would suffer a gravitational radiation
recoil.  A system in which a $10~\msun$ black hole merges into a
$130~\msun$ non-rotating black hole would have a recoil velocity
$20\kms \le v_{r} \le 200\kms$ \citep*{fhh04}.  Since $v_{r} \sim
\left(m_{1}/m_{0}\right)^{2}$, a merger between a $10~\msun$ black
hole and a seed black hole of mass of $250~\msun$, as discussed above,
would experience a recoil velocity $\la 50\kms$.  Mergers with lower
mass objects that are torn apart by tidal forces, such as white
dwarfs, would receive no gravitational radiation recoil.  Finally, a
range of interloper masses instead of the simplified single mass
population that we used here may also affect retention statistics
since a smaller interloper would impart smaller kicks while still
contributing to hardening.

Increasing the seed mass and the escape velocity will reduce the
number of field black holes ejected but not by enough.  As seen in
Figure~\ref{ejvm}, using a seed mass $m_{0} = 250~\msun$ and an escape
velocity $v_{\mathrm{esc}} = 70\kms$ reduces the number of black holes
ejected by 40\%, but this is still several factors more than are
available.  The Kozai-resonance-induced mergers and two-body captures,
however, are methods of merging without possibility of ejecting
stellar-mass black holes.  In order to reach our canonical
$1000~\msun$ intermediate mass while ejecting fewer than $10^{3}$
black holes, 70-80\% of the mergers must come from these ejectionless
methods.

\section{Implications for Gravitational Wave Detection}
\label{gwdetection}

\begin{figure}
\epsscale{0.75}
\rotatebox{90}{\plotone{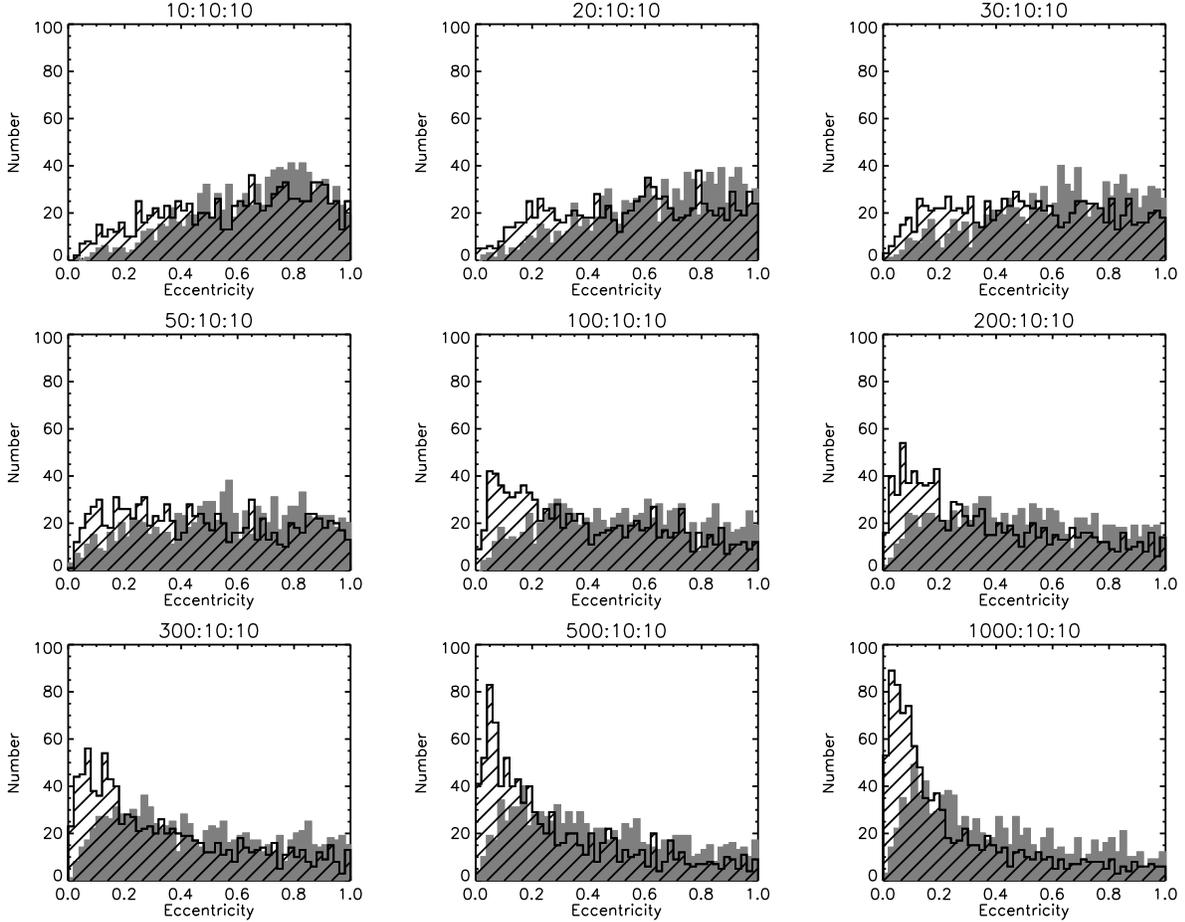}}
\caption{Distribution of eccentricities after integrating the Peters
  (1964) equations until in LISA band when the orbital frequency
  $\nu_{\mathrm{orb}} = 10^{-3}~\mathrm{Hz}$.  The solid histograms
  are the Newtonian only sequences, and the hatched histograms are
  sequences with gravitational radiation.  The sequences with
  gravitational radiation tend towards lower eccentricity since they
  have already started to circularize during the sequence.  There is
  more difference between the two cases at higher mass ratios since
  gravitational radiation is stronger.  Higher mass ratio binaries
  have lower eccentricities than lower mass ratio binaries since the
  latter start closer to merger after the final encounter.  The
  1000:10:10 mass ratio shows that a large number of detectable
  binaries would have $0.1 \la e \la 0.2$ such that they would likely
  be detectable by LISA with circular templates yet display measurable
  pericenter precession.}
\label{lisafreq}
\end{figure}

Our simulations make predictions interesting for gravitational wave
detection.  After the last encounter of a sequence, the binary will
merge due to gravitational radiation.  As the binary shrinks and
circularizes, the frequency of the gravitational radiation emitted
passes through the LISA band ($10^{-4}$ to $10^{0}~\mathrm{Hz}$)
\citep{d00} and then through the bands of ground-based detectors such
as LIGO, VIRGO, GEO-600, and TAMA ($10^{1}$ to $10^{3}~\mathrm{Hz}$)
\citep*{fetal97, s98, b00, aetal02}.  By the time the binaries are
detectable by ground-based instruments, they will have completely
circularized, but while in the LISA band, some will have measurable
eccentricities.  We calculate the distribution of eccentricities
detectable by LISA by integrating
Equations~\ref{petersa}~and~\ref{peterse} until the orbital frequency
reaches $\nu_{\mathrm{orb}} = 10^{-3}~\mathrm{Hz}$ at which point the
gravitational wave frequency is in LISA's most sensitive range and is
above the expected white dwarf background.  Figure~\ref{lisafreq}
shows the distribution of eccentricities for binaries with
gravitational radiation in the LISA band.  There are more low
eccentricities at higher mass ratios.  This is because at low mass
ratios each encounter takes a fractionally larger amount of energy
away from the binary than at high mass ratios.  Thus at low mass
ratios, the last encounter will tend to harden the binary such that it
is closer to merger.  At high mass ratios, however, encounters take a
smaller fractional amount of energy from the binary, and, thus, the
high mass ratio binaries have more time to circularize more during
their orbital decay.  For the 1000:10:10 mass ratio, a large fraction
of the eccentricities are in the range $0.1 \la e \la 0.2$ where the
binary is eccentric enough to display general relativistic effects
such as pericenter precession, but circular templates may be
sufficient for initial detection of the gravitational wave.  Finally,
because the first few hundred million years of a cluster's life
witness a large number of mergers, recently formed and nearby super
star clusters are promising sources of gravitational waves from IMBH
coalescence.

\section{Conclusions}
\label{conclusions}
We present results of numerical simulations of sequences of
binary-single black hole scattering events in a dense stellar
environment.  We simulate three-body encounters until the binary 
will merge due to gravitational radiation before the next encounter.  
In half of our simulations, we include the effect of gravitational 
radiation between encounters.

\emph{1. Sequences of high mass ratio encounters.}  Our simulations
cover a range of mass ratios including those corresponding to IMBHs in
stellar clusters.  Because the binaries simulated are tightly bound,
the encounters steadily shrink the binary's semimajor axis until it
merges.  The eccentricity, however, jumps chaotically between high and
low values over the course of a sequence.  Merger usually occurs at
high eccentricity since gravitational radiation is much stronger then.

\emph{2. Gravitational wave emission between encounters.}  The
inclusion of gravitational radiation between encounters affects the
simulations in several ways.  The extra source of shrinking caused by
gravitational wave emission has the effect of shortening the sequence
in terms of both the number of encounters and the total time, and the
circularization from gravitational waves has the effect of decreasing
the final eccentricity of the binary before it merges.

\emph{3. IMBH formation.}  Our simulations directly test the IMBH
formation model of \citet{mh02}.  We find that there is sufficient
time to build up to $1000~\msun$ when starting from $50~\msun$, but
our simulations also show that if there are a thousand $10~\msun$
black holes in the globular cluster, the seed black hole would only be
able to grow to $240~\msun$ before exhausting half of the black holes
in the cluster.  In addition, the probability of the binary's
remaining in the cluster during a growth from $50$ to $240~\msun$ is
small.  In order to avoid ejection from the cluster with a reasonable
probability, either the black hole must have a larger mass at the
onset of dynamical encounters, the cluster's escape velocity must be
larger, or the black hole must grow by some additional mechanisms such
as by Kozai-resonance-induced mergers, two-body captures, or from
smaller interlopers.

\emph{4. Gravitational wave detection.}  The mergers of binary black
hole systems are strong sources of detectable gravitational waves.  We
find that the merging binary will typically start with very high
eccentricity.  By the time the binary is detectable by the Advanced
LIGO detector, it will have completely circularized, but when
detectable by LISA, it may have moderate eccentricity ($0.1 \la e \la
0.2$) such that it will display general relativistic effects such as
pericenter precession and still possibly be detectable with circular
templates.  We find a high rate of mergers in the first few hundred
million years of a globular cluster.  This suggests that recently
formed, nearby super star clusters are promising sources for
gravitational radiation from IMBH coalescence.

Further work in this study will be to include a distribution of
interloper masses instead of a single population of $10~\msun$ black
holes.  A mass distribution of black holes is a more realistic model
of a cluster core and could change the outcomes of the sequences.
Exchanges will be more important since encounters with the more
prevalent smaller black holes may do most of the hardening until a 
more massive black hole exchanges into the binary.

\acknowledgements We thank J.~M. Fregeau, F.~A. Rasio, and
S.~Sigurdsson for helpful discussions and comments.  We are also
grateful for the hospitality of the Center for Gravitational Wave
Physics in which many fruitful ideas were born.  Many of the results
in this paper were obtained using the Beowulf cluster of the
University of Maryland department of astronomy.  This work was
supported in part by NASA grant NAG 5-13229.

\end{document}